# Linear and Non-Linear Landau Resonance of Kinetic Alfvén Waves: Consequences for Electron Distribution and Wave Spectrum in the Solar Wind

L. Rudakov[1], M. Mithaiwala, G. Ganguli and C. Crabtree

Plasma Physics Division, Naval Research Laboratory, Washington, DC 20375-5346
[1]Icarus Research Inc., P.O. Box 30780, Bethesda, MD 20824-0780 and
University of Maryand, Departments of Physics and Astronomy, College Park, Maryland 20742

## Abstract

Kinetic Alfven wave turbulence in solar wind is considered and it is shown that non-Maxwellian electron distribution function has a significant effect on the dynamics of the solar wind plasmas. Linear Landau damping leads to the formation of a plateau in the parallel electron distribution function which diminishes the Landau damping rate significantly. Nonlinear scattering of waves by plasma particles is generalized to short wavelengths and it is found that for the solar wind parameters this scattering is the dominant process as compared to three wave decay and coalescence in the wave vector range $1/\rho_i < k < \omega_{pe}/c$. Incorporation of these effects lead to the steepening of the wave spectrum between the inertial and the dissipation ranges with a spectral index between 2 and 3. This region can be labeled as the scattering range. Such steepening has been observed in the solar wind plasmas.

# 1. Introduction

Dispersive Alfvén waves have been considered as a possible candidate responsible for the solar wind turbulence.[1,2] In this scenario the turbulent cascade of Alfven waves transfer energy from the inertial range consisting of scales larger than the proton gyro-radius $\rho_i$ to scales smaller than the gyro-radius, $k_\perp \rho_i > 1$. Finally, the turbulence is damped at electron inertial scale $k_\perp c/\omega_{pe} \sim 1$ where $k_\perp$ is the perpendicular wave vector and $\omega_{pe}$ is the electron plasma frequency. At smaller scales the wave vector spectrum of the turbulence is highly anisotropic with energy concentrated in wave vectors nearly perpendicular to the mean magnetic field $B_0$ such that $k_\perp \gg k_\parallel$. These small scale $(k_\perp \rho_i > 1)$ waves have been identified as the Kinetic Alfvén waves (KAW). It should be noted that the frequency spectrum in the solar wind is unknown. Satellites measure the Doppler shifted frequency $\omega_k - k_\perp V_{SW}$, where $V_{SW}$ is the solar wind velocity whose Alfvén Mach number is in the range $5-7$ and $\omega_k$ is the wave frequency. Because $\omega_k \ll k_\perp V_{SW}$ satellites effectively measure the wave vector $k_\perp$.[3]

In the inertial range the observed turbulent spectrum closely follows the Kolmogorov scaling of $1/k^{5/3}$.[3,4] Similar to the scaling arguments of Kolmogorov, it was shown that in the electron-reduced-magneto-hydrodynamics model the turbulent cascade has a magnetic energy spectrum of $1/k_\perp^{7/3}$.[1,2] These general scaling arguments neglect the large Landau damping of KAW. However, the importance of electron Landau damping for KAW has been recognized.[5] Others have argued that the damping of KAW computed by assuming a Maxwellian distribution is too strong to create a steepened power spectrum but rather leads to sharp cutoffs.[6,7,8]

The general scaling arguments often used to describe the solar wind turbulence ignore the important phenomenon of induced non-linear scattering of KAW, also referred to as non-linear Landau damping. Using weak turbulence theory, Hasegawa and Chen found in 1976, that the rate of nonlinear (NL) scattering of the long wavelength dispersive Alfven waves, $k_\perp \rho_i \ll 1$, by plasma ions and mentioned that it "could be much larger than the rate of cascade evolution predicted by MHD theory."[9] The motivation of their work was to examine the plasma heating in Tokomaks using Alfven waves by including the NL scattering of the long wavelength in shorter wavelength for which the linear Landau damping is much larger. The decay and coalescence of long wavelength dispersive Alfven waves using the weak turbulence formalism as well as their parametric decay and interaction with convective cells has also been considered.[10,11,12].

The purpose of this article is to consider the induced NL wave-particle and wave-wave scattering of short wavelength KAW and assess their effects on the solar wind plasma turbulence. When considering the NL evolution of KAW it is necessary to simultaneously solve the quasi-linear equation for electrons and obtain the realistic electron distribution function which is modified due to Landau damping. Using a modified electron distribution function with a plateau in the parallel velocity distribution the NL scattering of KAW are generalized to short wavelengths $k_\perp^2 \rho_i^2 \gg 1$. The NL scattering of KAW by electrons leads to a steepened power law of the turbulent spectrum consistent with solar wind observations.

In solar wind the Landau damping time of the short wavelength KAW for a Maxwellian distribution function is few tenths of second while time of flight of solar wind to earth is a few days. Hence the electron distribution function can change quickly and affect the plasma dynamics during the lifetime of the solar wind plasma. The proton kinetic energy in fast solar wind is about a few KeV while electron and ion temperature is about a few tens eV and hence even if a tiny fraction of the fast solar wind energy is converted into KAW and subsequently Landau damped then the electron distribution can change significantly. Additionally, the Coulomb electron collision time for solar wind parameters[3] is about $10^5$ s and hence the electron distribution function will not be fully thermalized within the flight time from the Sun to the Earth. Therefore a Maxwellian distribution for electrons in the solar wind plasma cannot be justified.

After the plateau formation in the distribution there is no longer dissipation due to Landau damping for $k_\perp^2 c^2 / \omega_{pe}^2 \ll 1$, and therefore identifying the region $1/\rho_i^2 \ll k_\perp^2 \ll \omega_{pe}^2 / c^2$ as the "dissipation region" would be somewhat misleading. Recent observations in the solar wind indicate that there is an intermediate range between the inertial range $k_\perp^2 \rho_i^2 < 1$ and the onset of dissipation $k_\perp^2 c^2 / \omega_{pe}^2 \sim 1$.[3,4] From our study it seems appropriate to refer to this range as the "scattering range."

In section 2 a brief review of the linear theory of KAW is given. In section 3 it is shown that KAW turbulent spectrum evolution in the range $k_\perp^2 \rho_i^2 > 1$ can be treated using the "weak turbulence" theory. The importance of quasilinear relaxation of the parallel electron distribution function $f(v_\parallel)$ while evaluating linear and NL Landau damping in solar wind is emphasized. In section 4 the analysis of NL scattering of KAW by electrons is given. In section 5 we discuss the three wave decay and coalescence in NL interactions which can cascade the wave energy toward larger frequency and shorter scale. However, the calculated wave-wave scattering rate is found to be much smaller than the wave-particle scattering for parameters relevant to the solar wind turbulence in the range $k_\perp^2 c^2/\omega_{pe}^2 \ll 1$. Ultimately the coalescence of KAW at $k_\perp^2 c^2/\omega_{pe}^2 \sim 1$ transfers the energy from the scattering range into dissipation range $k_\perp^2 c^2/\omega_{pe}^2 > 1$.

## 2. Linear theory

We consider the short wavelength limit of Alfven wave branch, $k_\perp^2 \rho_i^2 = (k_\perp^2 c^2/\omega_{pi}^2)\beta_i > 1$. In a plasma with Maxwellian distribution for ions and electrons the dispersion relation for planar KAW $\vec{E} = \vec{E}_k \exp(-i\omega t + ik_x x + ik_z z)$ with $\omega \ll \Omega_i$ and $\omega/k_\parallel v_{te} < 1$ is (from Eq. (9), G. Howes, 2006)[13]

$$\omega^2 = \frac{k_\parallel^2 ((T_i + T_e)/M)}{1+(\beta_i+\beta_e)/2} \frac{k_\perp^2 c^2}{\omega_{pi}^2}\left(1 - i\sqrt{\pi}\frac{\omega}{k_\parallel v_{te}} T_e/(T_i+T_e)\right). \quad (1)$$

Here $k_\parallel \equiv k_z$, $k_\perp \equiv k_x$ are the wave vector components along and normal to the magnetic field $B_0$. KAW appears as elongated filaments along $B_0$ with $k_\parallel/k_\perp \sim (\omega/\Omega_i)/k_\perp^2 \rho_i^2 \ll 1$. The density is denoted by $n_s$ (where s = e or i denote electrons or ions respectively), $\beta_s \equiv 8\pi n_s T_s/B_o^2$, the temperature $T_s \equiv m_s v_{ts}^2/2$ for thermal speed $v_{ts}$, gyro-radius $\rho_s \equiv v_{ts}/\Omega_s$ with cyclotron frequency $\Omega_s \equiv eB_0/m_s c$, the Alfven speed $V_A = B_0/(4\pi n_i M)^{1/2}$, and the plasma frequency $\omega_{ps}^2 \equiv 4\pi n e^2/m_s$. The mass of species will be denoted by $m$ for electrons and $M$ for ions. In the solar wind $\beta_i \sim \beta_e \sim O(1)$.[3] For simplicity of calculations it is assumed that $m/M \ll (\beta_i, \beta_e) \ll 1$ but our analysis indicates that the estimates obtained under this assumption can be extended up to $\beta \sim 1$. A significant advantage of the $m/M \ll (\beta_i, \beta_e) \ll 1$ approximation is that the KAW is distinct from magnetosonic waves, which makes the analysis tractable.

The KAW is characterized by electric fields, $E_z$, $E_x$, vector potential $A_z$, and magnetic field, $B_y$. For waves with dispersion given in Eq. (1) the electric field components are

$$E_\parallel = -ik_\parallel \varphi - i\omega A_\parallel / c = -ik_\parallel \varphi + ik_\parallel \varphi (T_e + T_i)/T_i = ik_\parallel \varphi T_e / T_i,$$
$$E_\perp = -ik_\perp \varphi \tag{2}$$

The electro-magnetic (EM) component in $E_\parallel$ is not small and is in the opposite direction compared to the electro-static (ES) component, $-ik_\parallel \varphi$. The perpendicular electric field $E_\perp \gg E_\parallel$ and is mainly ES. EM component of $E_\perp$ is small, $\sim ik_\perp \varphi (k_\parallel^2 / k_\perp^2)$, and will be neglected. In this short wavelength field ions are effectively unmagnetized and their density perturbation can be described by the Boltzmann distribution, $\delta n_i / n = -e\varphi / T_i$. The ion Landau damping due to resonance $\omega = k_\parallel v_{\parallel i}$ is small as it is proportional to $\exp(-k_\perp^2 \rho_i^2 / \beta_i)$ and can be neglected. Magnetized electrons oscillate in $E_\parallel$ electric field Eq. (2) according to the equation $-T_e \nabla_\parallel \delta n_e - eE_\parallel n = 0$ and their density perturbation is $\delta n_e / n = -e\varphi / T_i$, which is same as for the ions. The magnetic field perturbation is

$$\frac{|\delta B_y|^2}{B_0^2} = \frac{\beta_i + \beta_e}{2} \frac{|e\varphi|^2}{T_i^2} = \frac{\beta_i + \beta_e}{2} \frac{|\delta n|^2}{n_0^2}. \tag{3}$$

KAW energy density is the sum of the ion kinetic energy, the electron kinetic energy, and the magnetic energy

$$W_k = \frac{n_0 T_i}{2}\frac{|\delta n|^2}{n_0^2} + \frac{n_0 T_e}{2}\frac{|\delta n|^2}{n_0^2} + \frac{|\delta B_y|^2}{8\pi} = \frac{|\delta B_y|^2}{4\pi}. \tag{4}$$

The electric field equations for KAW in the scattering range, $\beta_i M/m \gg k_\perp^2 \rho_i^2 \gg 1$, for arbitrary electron distribution function can be obtained from Ampere's law $\vec{\nabla}\times\vec{B} = 4\pi\vec{j}/c$ along with $\vec{B}=\vec{\nabla}\times\vec{A}$, $\vec{\nabla}\cdot\vec{A}=0$ and the potential equation $\vec{E} = i\omega\vec{A}/c + i\vec{k}\varphi$. The electron and ion current can be obtained from the susceptibility in a magnetized plasma.[14] Eliminating the vector-potential $\vec{A}$ from the time derivative of Ampere's law we obtain the equation which expresses the electric field vector $\vec{E}$ in terms of the electrostatic potential $\varphi$,

$$\bar{k}^2\left(\vec{E}+\vec{\nabla}\varphi\right) = (m/M)\left(-i\bar{\omega}(\vec{E}\times\vec{b}) + \frac{2\bar{\omega}^2 \vec{E}_x}{k_\perp^2\rho_i^2}\right) + \int\frac{\omega v_z dv_z dv_\perp^2}{\omega - k_z v_z}\frac{df_{0e}}{n_0 dv_z}\vec{E}_z. \tag{5}$$

Here $\bar{\omega} = \omega/\Omega_i$, $\bar{k}^2 = k^2 c^2/\omega_{pe}^2$, $\bar{k}^2 = \bar{k}_z^2 + \bar{k}_y^2$. The z-component of Eq. (5) yields

$$E_z = -ik_z\varphi\frac{\bar{k}^2}{\left(\bar{k}^2 - \int\frac{\omega v_z dv_z dv_\perp^2}{\omega-k_z v_z}\frac{df_{0e}}{n_0 dv_z}\right)}. \tag{6}$$

The x-component of Eq. (5) yields

$$E_x = -ik_x\varphi\left(1 + O(k_z^2/k_x^2)\right). \tag{7}$$

The dispersion relation for KAW follows by setting the divergence of Eq. (5) [Eq. (5) represents the time derivative of the total current] to zero along with $\varphi \neq 0$ and the electric field components given by Eqs. (6), (7).

$$\omega^2 \left\{ 1 - \frac{\bar{k}^2}{\int \frac{\omega v_z dv_z dv_\perp^2}{\omega - k_z v_z} \frac{df_{0e}}{n_0 dv_z}} \right\} = k_z^2 \bar{k}_x^2 T_i / m \qquad (8)$$

## 3. Quasi-linear evolution of electron distribution function

Measurements of the magnetic field spectrum of solar wind turbulence show that for $k_\perp \rho_i > 1$ the magnetic field oscillation $\delta B / B_0$ is small.[3] The magnetic field energy $(\delta B)^2 \sim (\delta B_k)^2 k_\perp$ shows a three order of magnitude drop between $10^{-3} \leq k_\perp \rho_i \leq 1$ (see Fig. 1). It is clear that in the range $k_\perp \rho_i > 1$ we deal with the case $\delta B / B_0 \sim \delta n / n_0 < 0.03$ and it is possible to expand the Vlasov equation over this small parameter and to use the ideas and techniques of weak turbulence theory.

The weak turbulence of the collisionless plasma is based on the key assumption that there exists an ensemble of particles given by Vlasov equation and an ensemble of waves or plasmons with uncorrelated phases.[15,16,17] Plasmons and particles interact due to linear Landau or cyclotron resonance $\omega_k - k_z v_z - l\Omega_s = 0$ and non-linear (NL) Landau resonance of beat waves $(\omega_{k2} - \omega_{k1}) - (k_{2z} - k_{1z}) v_z = 0$, which gives NL waves scattering. Linear damping (growth) should be small for NL scattering to compete. Also plasmons may interact in three-wave resonance $\omega_{k1} \pm \omega_{k2} = \omega_{k3}, \vec{k}_1 \pm \vec{k}_2 = \vec{k}_3$. However it will be shown that the dominant nonlinear effect for KAW in the range $k^2 c^2 / \omega_{pe}^2 \ll 1$ in solar wind is NL scattering of waves by plasma particles.

In weak turbulence a given plasmon interacts with a particle or waves independently from other particles or waves. So, we can calculate the linear or NL (parametric) interaction of plasmons and then sum them over the turbulence bandwidth to obtain the scattering rate in turbulence. Below supercripts $^{(1), (2), (3)}$ correspond to the first, second and third order perturbations. For the first order perturbed electric field the superscript will be omitted, i.e. $\vec{E} = 2^{-1/2}\{\vec{E}_k \exp(-i\omega t + ik_x x + ik_z z + i\phi_k) + cc\}$, and $\phi_k$ is the wave phase, $<\phi_{k1}\phi_{k2}> = \delta_{k1-k2}$. For calculating NL effects we will use Vlasov equation in drift approximation[18,9],

$$\frac{\partial f}{\partial t} + \vec{\nabla}(\vec{b}v_z f) + \vec{\nabla}\left(c\frac{\vec{E}\times\vec{b}}{B_0}f\right) - \frac{e}{m}E_z\frac{\partial f(x,z,v_z,\mu)}{\partial v_z} = 0. \quad (9)$$

Note that the magnetic moment of electrons $\mu = mv_\perp^2/B$ is the adiabatic invariant, i.e., $d\mu/dt = 0$. In addition, $\delta B/B_0$ is small. Hence, there is no change in the $v_\perp$ distribution. However, the parallel distribution function can change. In the following we take this into account and obtain the appropriate parallel electron distribution function. Linearizing Eq. (9) we get

$$f_k^{(1)} = \frac{i}{(\omega_k - k_z v_z)}\frac{eE_{zk}}{m}\frac{\partial f_0(v_z, v_\perp^2)}{\partial v_z} \quad (10)$$

Substituting Eq. (10) in Eq. (9) and averaging over phases $\phi_k$ we obtain the quasi-linear equation[15,16,17] which describes the evolution of the slowly varying distribution function in time $f_0(t, v_z, v_\perp^2)$

$$\frac{\partial f_0}{\partial t} = \pi\frac{\partial}{\partial v_z}\int\frac{e^2}{2m^2}\sum_k |E_{zk}|^2 \delta(\omega_k - k_z v_z)dk_\parallel \frac{\partial f_0}{\partial v_z}. \quad (11)$$

Note that electron heating given by Eq. (11) is irreversible because the plasma entropy increases.

$$S = -\int f_e \ln f_e dv_z dv_\perp^2 , \quad \frac{\partial S}{\partial t} = \pi \int dv_z dv_\perp^2 \int \frac{e^2}{2m^2} \sum_k |E_{zk}|^2 \delta(\omega_k - k_z v_z) dk_z \frac{1}{f_e}\left(\frac{\partial f_e}{\partial v_z}\right)^2 > 0 \quad (12)$$

We start with a Maxwellian electron distribution function with thermal velocity $v_{te}^2 = 2T_e/m$. Quasi-linear diffusion establishes and maintains a plateau in electron distribution in the velocity range $V_A < v_z < v_m$, where $v_m$ is the maximum velocity in the plateau region (see Fig. 2). For the interval $v_z < V_A = B_0/(4\pi n_0 M)^{1/2}$ there are no waves with appropriate phase velocity for resonance. Based on these considerations and for simplicity we approximate the time asymptotic solution of Eq. (11) as (see Fig. 2),

$$\begin{aligned}
f_{0e}(v_\parallel, v_\perp)/n_0 &= (\alpha/\pi^{1/2} v_{te}^2 v_{tc})\exp(-v_\parallel^2/v_{tc}^2 - v_\perp^2/v_{te}^2) \\
&+ \exp(-v_\perp^2/v_{te}^2)\theta(v_m^2 - v_\parallel^2)/2v_m v_{te}^2 \\
&+ \exp(-(v^2 - v_m^2)/v_{te}^2)\theta(v_\parallel^2 - v_m^2)/2v_m v_{te}^2
\end{aligned} \quad (13)$$

where $\theta(v_m^2 - v_z^2)$ is the step function of amplitude one, $v_{tc} < V_A$ and $\alpha$ are parameters of the "cold" core, such that $\alpha \sim V_A/v_{te}$ is a small fraction of the total electron density. The core can evolve slowly due to NL Landau resonance if $(\omega_{k2} - \omega_{k1}) \leq (k_{2z} - k_{1z})v_{tc}$.

Now we recalculate the dispersion relation and $E_z$ using the modified distribution function Eq. (13), under the assumption that $v_{tc} < V_A < \omega/k_z < v_m$, to obtain the KAW dispersion relation,

$$\omega^2 = k_z^2(v_m^2 + T_i/m)\bar{k}^2/(1+\bar{k}^2), \quad E_z = ik_z\varphi(mv_m^2 - \bar{k}^2 T_i)/T_i(1+\bar{k}^2). \quad (14)$$

To satisfy the inequality $\omega/k_z < v_m$ we should keep $\bar{k}^2 < mv_m^2/T_i$. For established plateau in the electron distribution function (e.g., Eq. (13)) the Landau damping is minimized and NL scattering of KAW can easily dominate over linear damping.

## 4. NL scattering by plasma particles

The linear dispersion relation (14) was obtained from the fourth term on the left hand side of the Vlasov equation (9) because $E_\perp = -ik_\perp \varphi$ and the third term $\vec{\nabla}(c\vec{E}_\perp \times \vec{b}/B_0)f_e^0 = 0$. However the NL terms of the wave turbulence equations, arising from $f_k^{(2)}$ and $f_k^{(3)}$, are obtained from the third term since $(c\vec{E} \times \vec{b}/B_0)\vec{\nabla} f_e^{(1),(2)} \neq 0$ since the fourth term is smaller by $k_\parallel / k_\perp$. Thus it is only necessary to use a simplified version of the drift kinetic equation,

$$\frac{\partial f}{\partial t} + v_z \frac{\partial f}{\partial z} + c\frac{\vec{E} \times \vec{b}}{B_0} \vec{\nabla} f = 0. \tag{15}$$

To calculate NL scattering rate, the second order slow electron density perturbation due to beat waves with frequency $\omega_{k2} - \omega_{k1}$ is needed. The second order distribution function, after iterating (15) is

$$f_{k2-k1}^{(2)} = \frac{1}{(\omega_{k2} - \omega_{k1}) - (k_{2z} - k_{1z})v_z} \left( \begin{array}{c} -\frac{i(\vec{k}_1 \times \vec{k}_2)\vec{b}}{\Omega_e} \frac{mv_m^2}{T_i} \frac{e^2 \varphi_{k1}^* \varphi_{k2}}{m^2} \\ \times \left( \frac{k_{2z}}{(\omega_{k2} - k_{2z}v_z)} - \frac{k_{1z}}{(\omega_{k1} - k_{1z}v_z)} \right) + i\frac{eE_{z,k2-k1}^{(2)}}{m} \end{array} \right) \frac{\partial f_0(v_z)}{\partial v_z} \tag{16}$$

The second order electric field $E_z^{(2)}$ maintains quasi-neutrality. From the electric field Eq. (6),

$$E_{z,k2-k1}^{(2)} = -\frac{i(k_{2z} - k_{1z})\varphi_{k2-k1}^{(2)}(\vec{\bar{k}}_2 - \vec{\bar{k}}_1)^2}{(\vec{\bar{k}}_2 - \vec{\bar{k}}_1)^2 - \frac{(\omega_{k2} - \omega_{k1})^2}{(k_{2z} - k_{1z})^2 v_m^2 - (\omega_{k2} - \omega_{k1})^2}}. \tag{17}$$

For NL scattering by cold electrons, $(\omega_{k2} - \omega_{k1})^2/(k_{2z} - k_{1z})^2 < v_{tc}^2 \ll v_m^2$, and for $(\vec{k}_2 - \vec{k}_1)^2 \rho_i^2 \gg 1$ the electric field is quasi-electrostatic, $E_{z,k2-k1}^{(2)} = -i(k_{2z} - k_{1z})\varphi_{k2-k1}^{(2)}$. The third order electron distribution function is

$$f_{k2}^{(3)} = \frac{-i}{(\omega_{k2} - k_{2z}v_z)} c \frac{(\vec{k}_1 \times \vec{k}_2)\vec{b}}{B_0} \left( \varphi_{k1} f_{k2-k1}^{(2)} - f_{k1} \varphi_{k2-k1}^{(2)} \right). \tag{18}$$

The potential $\varphi_{k2-k1}^{(2)}$ can be eliminated by using plasma quasi-neutrality on the slow-time scale $1/(\omega_{k2} - \omega_{k1})$

$$\delta n_{e,k2-k1}^{(2)} = \delta n_{i,k2-k1}^{(2)}. \tag{19}$$

The NL scattering rate follows from imaginary part of the equation of the plasma quasi-neutrality on the fast–time scale $1/\omega_{k2}$, $\delta n_{e,k2}^{(1)} + \delta n_{e,k2}^{(3)} = \delta n_{i,k2}^{(1)} + \delta n_{i,k2}^{(3)}$,

$$\gamma_{NL} \frac{\partial}{\partial \omega} \frac{\delta n_{e,k2}^{(1)}}{n_0} = \gamma_{NL} \frac{e\varphi_{k2}}{T_i} \frac{2(mv_m^2 + T_i)}{\omega T_i} = -\mathrm{Im} \frac{\delta n_{e,k2}^{(3)}}{n_0}. \tag{20}$$

We calculate $\delta n_e^{(2)}$ and $\delta n_e^{(3)}$ with the assumption that $(\omega_{k2} - \omega_{k1}) \sim k_z v_{tc} \ll \omega_{k2}$.

$$\frac{\delta n_e^{(2)}}{n_0} = \frac{-i\omega_{pe}^2}{\Omega_e \omega_{k1}} \frac{(\vec{k}_1 \times \vec{k}_2)\vec{b}\, \varphi_{k1}^* \varphi_{k2}}{4\pi n_0 T_i} \left( \frac{\alpha m v_m^2 (1 + \zeta_{ec} Z(\zeta_{ec}))}{T_c} + \frac{\omega_{k1}\omega_{k2}}{k_{1z}k_{2z}v_m^2} \right)$$
$$+ \frac{e\varphi_{k2-k1}^{(2)}}{mv_m^2} \left( 1 + \frac{\alpha m v_m^2 (1 + \zeta_{ec} Z(\zeta_{ec}))}{T_c} \right), \tag{21}$$

where $Z(\zeta_{ec})$ is the plasma dispersion function with argument

$\zeta_{ec} = (\omega_{k1} - \omega_{k2})/|k_{1\parallel} - k_{2\parallel}|v_{tc}$. Let $\bar{k}^2 = k^2 c^2/\omega_{pe}^2 \ll 1$ and $v_{tc}^2 \ll \omega^2/k_z^2 \ll v_m^2$,

$E_z = ik_z \varphi m v_m^2/T_i$. The ion density perturbation $\delta n_i^{(2)}$ in the short wavelength

$(\vec{k}_1 - \vec{k}_2)^2 \rho_i^2 \gg 1$ electrostatic electric field of the beat wave $\vec{E}_{\perp,k2-k1}^{(2)} = -i(\vec{k}_{2\perp} - \vec{k}_{1\perp})\varphi_{k2-k1}^{(2)}$

is[14]

$$\frac{\delta n_i^{(2)}}{n_0} = -\frac{e}{T_i}\varphi_{k2-k1}^{(2)}\left(1+\frac{(1+\zeta_i Z(\zeta_i))}{\pi^{1/2}k_\perp \rho_i}\right), \quad \zeta_i = \frac{(\omega_{k1}-\omega_{k2})}{|k_{1\|}-k_{2\|}|v_{ti}}. \tag{22}$$

Here we keep the small but important last term which accounts for ion oscillation along the magnetic field $B_0$ and Landau resonance with beat waves $(\omega_{k1}-\omega_{k2}) = (k_{1\|}-k_{2\|})v_{\|i}$.

Using plasma quasi-neutrality $\delta n_{e,k2-k1}^{(2)} = \delta n_{i,k2-k1}^{(2)}$ we calculate $\varphi_{k2-k1}^{(2)}$

$$\frac{e\varphi_{k2-k1}^{(2)}}{T_i} = \left(\frac{i\omega_{pe}^2}{\Omega_e \omega_{k1}}\frac{(\vec{k}_1\times\vec{k}_2)\vec{b}\,\varphi_{k1}^*\varphi_{k2}}{4\pi n_0 T_i}\right)\left(\frac{\alpha m v_m^2(1+\zeta_{ec}Z(\zeta_{ec}))}{T_c}+\frac{\omega_{k1}\omega_{k2}}{k_{1z}k_{2z}v_m^2}\right)\frac{mv_m^2}{mv_m^2+T_i}$$
$$\times\left\{1-\left(\frac{(1+\zeta_i Z(\zeta_i))}{\pi^{1/2}k_\perp\rho_i}+\frac{\alpha(1+\zeta_{ec}Z(\zeta_{ec}))T_i}{T_c}\right)\frac{mv_m^2}{mv_m^2+T_i}\right\}. \tag{23}$$

Calculating $\delta n_e^{(3)}$ by using Eq. (18) we obtain,

$$\frac{\delta n_{k2}^{(3)}}{n_0} = -c\frac{i\varphi_{k1}(\vec{k}_1\times\vec{k}_2)\vec{b}}{B_0\omega_{k2}}\frac{mv_m^2}{T_i}\left\{\begin{array}{l}-\dfrac{i\omega_{pe}^2}{\Omega_e\omega_{k2}}\dfrac{(\vec{k}_1\times\vec{k}_2)\vec{b}\,\varphi_{k1}^*\varphi_{k2}}{4\pi n_0 T_i}\left(\dfrac{\alpha mv_m^2(1+\zeta_{ec}Z(\zeta_{ec}))}{T_c}-\dfrac{\omega_{k1}\omega_{k2}}{k_{1z}k_{2z}v_m^2}\right)\\+\dfrac{e\varphi_{k2-k1}^{(2)}}{mv_m^2}\left(\dfrac{\alpha mv_m^2(1+\zeta_{ec}Z(\zeta_{ec}))}{T_c}+\dfrac{\omega_{k1}\omega_{k2}}{k_{1z}k_{2z}v_m^2}\right)\end{array}\right\}. \tag{24}$$

Substituting Eq. (23) in Eq. (24) we get

$$\frac{\delta n_{k2}^{(3)}}{n_0} = -\frac{e\varphi_{k2}}{T_i}\frac{c^2|\varphi_{k1}|^2(\vec{k}_1\times\vec{k}_2)_\|^2}{B_0^2\omega_{k2}^2}\frac{mv_m^2}{T_i}\left\{\begin{array}{l}\left(\dfrac{\alpha mv_m^2(1+\zeta_{ec}Z(\zeta_{ec}))}{T_c}-\dfrac{\omega_{k1}\omega_{k2}}{k_{1z}k_{2z}v_m^2}\right)\\-\left(\dfrac{\alpha mv_m^2(1+\zeta_{ec}Z(\zeta_{ec}))}{T_c}+\dfrac{\omega_{k1}\omega_{k2}}{k_{1z}k_{2z}v_m^2}\right)^2\dfrac{mv_m^2}{mv_m^2+T_i}\\\times\left\{1-\left(\dfrac{(1+\zeta_i Z(\zeta_i))}{\pi^{1/2}k_\perp\rho_i}+\dfrac{\alpha(1+\zeta_{ec}Z(\zeta_{ec}))T_i}{T_c}\right)\dfrac{mv_m^2}{mv_m^2+T_i}\right\}\end{array}\right\}. \tag{25}$$

Using Eq. (20) and Eq. (25) we get the nonlinear growth rate

$$\frac{\gamma_{NL}^{e,i}}{\omega} \sim \sum_{k1}\frac{c^2|\varphi_{k1}|^2(\vec{k}_1\times\vec{k}_2)_\|^2}{B_0^2\omega_{k2}^2}\frac{mv_m^2}{T_i}\left\{\frac{\alpha mv_m^2\zeta_{ec}\,\mathrm{Im}\,Z(\zeta_{ec})}{T_c}+\left(\frac{\omega_{k1}\omega_{k2}}{k_{1z}k_{2z}v_m^2}\frac{mv_m^2}{mv_m^2+T_i}\right)^2\frac{\zeta_i\,\mathrm{Im}\,Z(\zeta_i)}{\pi^{1/2}k_\perp\rho_i}\right\}. \tag{26}$$
$$\sim \sum_{k1}\frac{\Omega_i^2}{\omega_{k2}^2}\frac{|\delta B_{k1}|^2}{n_0 T_i}(\vec{k}_1\times\vec{k}_2)_\|^2\rho_i^4\frac{mv_m^2}{T_i}\left\{\frac{\alpha mv_m^2\zeta_{ec}\,\mathrm{Im}\,Z(\zeta_{ec})}{T_c}+\bar{k}_1^2\bar{k}_2^2\frac{\zeta_i\,\mathrm{Im}\,Z(\zeta_i)}{\pi^{1/2}k_\perp\rho_i}\right\}$$

The plasmon energy Eq. (4) is used in the last line to express NL scattering rate in terms of the magnetic field fluctuations $|\delta B_{k1}|^2$. The NL rate was calculated with the assumption that $\bar{k}^2 = k^2 c^2/\omega_{pe}^2 \ll 1$ and $v_{tc}^2 \ll \omega^2/k_z^2 \ll v_m^2$. Only the terms proportional to the first order in the small parameters $\alpha$ and $1/k_\perp \rho_i$ are retained.

In space experiments[3-5] the magnetic field fluctuations $|\delta B_k|^2$ are measured as a continuous distribution of the wave vector $k_\perp$ (see Fig. 1). In order to represent the scattering rate (26) as an experimentally meaningful quantity, the simplifying assumption that the KAW spectral wave energy $W_{k1} = N_{k1}\omega_{k1}$ is a function of $\omega_{k1}$ and $k_{1\perp}$ separately is introduced, i.e. $N_{k1} = N_{\omega 1} P(k_{1\perp})/P_0$, where $N_{k1}$ is "the number of plasmons", $P(k_{1\perp})$ and $N_{\omega 1}$ are the $k_{1\perp}$ and $\omega_{k1}$ distributions of plasmons and $P_0$ is a normalization factor[3]. Then the summation over $\vec{k}_1$ can be converted into an integral over $k_{1\perp} dk_{1\perp} dk_{1\parallel}$. Using the KAW dispersion relation (14), $dk_\parallel \sim d\omega/k_\perp$, the scattering rate (26) can be rewritten as,

$$\gamma_{NL}^e \sim \sum_{k1} A_{k1,k2} N_{k1} \zeta_{ec,k2-k1} \operatorname{Im} Z_{ec,k2-k1} \to \int A_{k1,k2} N_{\omega 1} \zeta_{ec,k2-k1} \operatorname{Im} Z_{ec,k2-k1} d\omega_{k1} \frac{P(k_{1\perp})}{P_0} dk_{1\perp}$$
$$= \int A_{k1,k2} (N_{\omega 1} + (\omega_{k1} - \omega_{k1}) dN_{\omega 1}/d\omega_{k1}) \zeta_{ec,k2-k1} \operatorname{Im} Z_{ec,k-k1} d\omega_{k1} \frac{P(k_{1\perp})}{P_0} dk_{1\perp}$$
$$\sim \int A_{k1,k2} \left( N_\omega + \frac{v_{ec}(\bar{k}_{2\perp} - \bar{k}_{1\perp})}{v_m \bar{k}_{2\perp} \bar{k}_{1\perp}} \omega \frac{dN_\omega}{d\omega} \right) \frac{v_{ec}(\bar{k}_{2\perp} - \bar{k}_{1\perp})}{v_m \bar{k}_{2\perp} \bar{k}_{1\perp}} \frac{P(k_{1\perp})}{P_0} dk_{1\perp}$$

(27)

Finally, for the case $k_{1\perp}^2 \rho_i^2 > k_{2\perp}^2 \rho_i^2 \gg 1$ and $T_i \sim m v_m^2$, the estimate of the NL scattering rate in $\vec{k}_\perp$-space by cold electrons, represented by the first term in the bracket of Eq. (27) is,

$$\gamma_{NL}^{k1\to k2} \sim \alpha\left(\frac{v_m}{v_{tc}}\right)\frac{\Omega_i^2}{\beta_i \delta\omega}\left(\frac{|\delta B_\omega|^2}{B_0^2}\right)\int (\vec{k}_{1\perp}\times\vec{k}_{2\perp})_\parallel^2 \rho_i^4 \frac{(\bar{k}_{2\perp}-\bar{k}_{1\perp})}{\bar{k}_{2\perp}\bar{k}_{1\perp}}\frac{P(k_{1\perp})}{P_0}dk_{1\perp} \ . \tag{28}$$

Here $\delta\omega$ is the width of the spectral distribution over frequency. The rate of the inverse cascade in the frequency space due to NL scattering by cold electrons, corresponding to the second term in the bracket of Eq. (27) is

$$\gamma_{NL}^{\omega 1\to \omega 2} \sim \frac{\alpha\Omega_i^2}{\beta_i}\frac{d}{d\omega}\left(\frac{|\delta B_\omega|^2}{B_0^2}\right)\int (\vec{k}_{1\perp}\times\vec{k}_{2\perp})_\parallel^2 \rho_i^4 \left(\frac{\bar{k}_{2\perp}-\bar{k}_{1\perp}}{\bar{k}_{2\perp}\bar{k}_{1\perp}}\right)^2 \frac{P(k_{1\perp})}{P_0}dk_{1\perp} \ . \tag{29}$$

In each scattering event the wave vector value and direction changes strongly while the frequency decreases by small steps, $(\omega_{k1}-\omega_{k2})/\omega_{k1} \sim (v_{tc}/v_m)(\bar{k}_{2\perp}-\bar{k}_{1\perp})/\bar{k}_{1\perp}\bar{k}_{2\perp}$.

## 5. Three wave resonance

In addition to NL scattering, there is possibility for three-wave interaction of KAW as well if the resonance conditions can be satisfied,

$$\omega_{k1}=\omega_{k2}\pm\omega_{k3}, \ \vec{k}_1=\vec{k}_2\pm\vec{k}_3. \tag{30}$$

The three-wave interaction of dispersive MHD waves for $k_\perp^2 \rho_i^2 \ll 1$ in the weak turbulence limit has been analyzed.[10,11] Here we investigate the solar wind KAW decay in the short wavelength range $k_\perp^2 \rho_i^2 \gg 1$. For example, the resonance conditions can be met with a wave $\omega_{k1}, k_{1z}, \vec{k}_{1\perp}$ and two decay waves $\omega_{k2}=\omega_{k3}=\omega_{k1}/2$, $k_{2z}=k_{3z}=k_{1z}/2$, when the vectors $\vec{k}_{2\perp}$ and $\vec{k}_{3\perp}$ are in the same plane as $\vec{k}_{1\perp}$ but both inclined by $\pm 60^o$ relative to $\vec{k}_{1\perp}$ and all $\vec{k}_\perp$ vectors have the same magnitude. In the solar wind plasma where $T_e \sim T_i$ and $\beta_e \sim \beta_i \sim 1$ low frequency, $\omega \ll \Omega_i$, ion sound waves and magnetosonic waves in the wave vector range $k_\perp^2 \rho_i^2 \gg 1$ do not exist.   We recalculate $\delta n_e^{(2)}$ and $\delta n_e^{(3)}$ under the assumption that $(\omega_{k2} - \omega_{k1}) \sim \omega_{k2}$ neglecting NL Landau resonance with "cold core" electrons. Let $\bar{k}^2 = k^2 c^2 / \omega_{pe}^2 << 1$, $\omega^2 / k_z^2 << v_m^2$, $E_z = ik_z \varphi m v_m^2 / T_i$, $m / \beta_i M << (\bar{\vec{k}}_2 - \bar{\vec{k}}_1)^2 << 1$, $(\omega_{k2} - \omega_{k1})^2 << (k_{2z} - k_{1z})^2 v_m^2$, and $E_{z,k2-k1}^{(2)}$ is given by Eq. (17).

$$\frac{\delta n_e^{(2)}}{n_0} = \frac{-i\omega_{pe}^2}{\Omega_e \omega_{k1}} \frac{(\vec{k}_1 \times \vec{k}_2)\vec{b}\, \varphi_{k1}^* \varphi_{k2}}{4\pi n_0 T_i} \frac{\omega_{k1}\omega_{k2}}{k_{1z}k_{2z}v_m^2} + i\frac{eE_{z,k2-k1}^{(2)}}{(k_{2z}-k_{1z})mv_m^2}$$

$$= \frac{-i\omega_{pe}^2}{\Omega_e \omega_{k1}} \frac{(\vec{k}_1 \times \vec{k}_2)\vec{b}\, \varphi_{k1}^* \varphi_{k2}}{4\pi n_0 T_i} \frac{\omega_{k1}\omega_{k2}}{k_{1z}k_{2z}v_m^2} + \frac{(\bar{\vec{k}}_2 - \bar{\vec{k}}_1)^2 (k_{2z}-k_{1z})^2}{(\bar{\vec{k}}_2 - \bar{\vec{k}}_1)^2 (k_{2z}-k_{1z})^2 v_m^2 - (\omega_{k2}-\omega_{k1})^2} \frac{e\varphi_{k2-k1}^{(2)}}{m} \quad (31)$$

The ion density perturbation $\delta n_i^{(2)}$ in the short wavelength $(\vec{k}_{2\perp} - \vec{k}_{1\perp})^2 \rho_i^2 >> 1$ beat wave $\vec{E}_{\perp,k2-k1}^{(2)} = -i(\vec{k}_{2\perp} - \vec{k}_{1\perp})\varphi_{k2-k1}^{(2)}$ is

$$\frac{\delta n_i^{(2)}}{n_0} = -\frac{e}{T_i} \varphi_{k2-k1}^{(2)} \quad . \quad (32)$$

Using plasma quasi-neutrality in slow motion $\delta n_{e,k2-k1}^{(2)} = \delta n_{i,k2-k1}^{(2)}$ we calculate $\varphi_{k2-k1}^{(2)}$

$$\frac{e\varphi_{k2-k1}^{(2)}}{T_i} = \left(\frac{i\omega_{pe}^2}{\Omega_e \omega_{k1}} \frac{(\vec{k}_1 \times \vec{k}_2)\vec{b}\, \varphi_{k1}^* \varphi_{k2}}{4\pi n_0 T_i}\right) \frac{\omega_{k1}\omega_{k2}}{k_{1z}k_{2z}v_m^2} \frac{(\bar{\vec{k}}_2 - \bar{\vec{k}}_1)^2 (k_{2z}-k_{1z})^2 v_m^2 - (\omega_{k2}-\omega_{k1})^2}{(\bar{\vec{k}}_2 - \bar{\vec{k}}_1)^2 (k_{2z}-k_{1z})^2 (T_i/m + v_m^2) - (\omega_{k2}-\omega_{k1})^2} \quad (33)$$

Now we calculate $\delta n_e^{(3)}$ using Eq. (18)

$$\frac{\delta n_{k2}^{(3)}}{n_0} = -c\frac{i\varphi_{k1}(\vec{k}_1 \times \vec{k}_2)\vec{b}}{B_0 \omega_{k2}} \frac{mv_m^2}{T_i} \frac{\omega_{k1}\omega_{k2}}{k_{1z}k_{2z}v_m^2} \left\{\frac{i\omega_{pe}^2}{\Omega_e \omega_{k2}} \frac{(\vec{k}_1 \times \vec{k}_2)\vec{b}\, \varphi_{k1}^* \varphi_{k2}}{4\pi n_0 T_i} + \frac{e\varphi_{k2-k1}^{(2)}}{mv_m^2}\right\} \quad . \quad (34)$$

Substituting Eq. (33) in Eq. (34) we get

$$\frac{\delta n_{k2}^{(3)}}{n_0} = \frac{e\varphi_{k2}}{T_i} \frac{c^2 |\varphi_{k1}|^2 (\vec{k}_1 \times \vec{k}_2)_\parallel^2}{B_0^2 \omega_{k2}^2} \frac{mv_m^2}{T_i} \left(\frac{\omega_{k1}\omega_{k2}}{k_{1z}k_{2z}v_m^2}\right)$$
$$\times \left\{1 + \left(\frac{\omega_{k1}\omega_{k2}}{k_{1z}k_{2z}v_m^2}\right) \frac{(\bar{\vec{k}}_2 - \bar{\vec{k}}_1)^2 (k_{2z}-k_{1z})^2 v_m^2 - (\omega_{k2}-\omega_{k1})^2}{(\bar{\vec{k}}_2 - \bar{\vec{k}}_1)^2 (k_{2z}-k_{1z})^2 (T_i/m + v_m^2) - (\omega_{k2}-\omega_{k1})^2}\right\} \quad . \quad (35)$$

Now using Eq. (20) and Eq. (35) we get

$$\frac{\gamma_{NL}}{\omega} \sim \sum_{k1} \frac{c^2 |\varphi_{k1}|^2 (\vec{k}_1 \times \vec{k}_2)_{\parallel}^2}{B_0^2 \omega_{k2}^2} \frac{mv_m^2}{T_i} \left(\frac{\omega_{k1}\omega_{k2}}{k_{1z}k_{2z}v_m^2}\right)^2$$
$$\times \text{Im}\left\{\frac{(\overline{\vec{k}}_2 - \overline{\vec{k}}_1)^2 (k_{2z} - k_{1z})^2 v_m^2 - (\omega_{k2} - \omega_{k1})^2}{(\overline{\vec{k}}_2 - \overline{\vec{k}}_1)^2 (k_{2z} - k_{1z})^2 (T_i/m + v_m^2) - (\omega_{k2} - \omega_{k1})^2}\right\}. \qquad (36)$$

The denominator of Eq. (36) contains the decay condition, Eq. (30). Eq. (36) is general and holds for both narrow and broadband spectra. For a narrowband, $\delta\omega \ll \gamma_{NL}$, spectra of mother waves the parametric decay of KAW can be obtained from Eq. (36) in the limit $mv_m^2 \sim T_i$ as,

$$\gamma_{NL}^2 \sim \frac{c^2 |\varphi_{k1}|^2 (\vec{k}_1 \times \vec{k}_2)_{\parallel}^2}{B_0^2} \left(\frac{\omega_{k1}\omega_{k2}}{k_{1z}k_{2z}v_m^2}\right)^2 \sim \frac{\Omega_i^2}{\omega_{k2}^2} \frac{|\delta B_{k1}|^2}{B_0^2} \frac{(\vec{k}_1 \times \vec{k}_2)_{\parallel}^2 \rho_i^4}{(\beta_i + \beta_e)} \frac{\overline{k}_1^2 \overline{k}_2^2}{(1+\overline{k}_1^2)(1+\overline{k}_2^2)}. \qquad (37)$$

Note that the NL rates given in Eqs. (36) and (37) are derived using the modified electron distribution function as given in Eq. (13). Similar calculation of the rate of parametric decay can be done for a Maxwellian distribution of electrons. But because large linear Landau damping of the short wavelength KAW for both mother and daughter waves ($\gamma_{k1}, \gamma_{k2}$) the threshold for parametric decay is large. In this case instead of Eq. (37) we have $(\gamma + \gamma_{k1})(\gamma + \gamma_{k2}) = \gamma_{NL}^2$ with threshold $\gamma_{NL}^2 > \gamma_{k1}\gamma_{k2}$.

For broadband spectra of mother waves, $\delta\omega \gg \gamma_{NL}$, Eq. (36) leads to the rate

$$\gamma_{NL}^{3\omega} \sim \sum_{k1} \frac{c^2 |\varphi_{k1}|^2 (\vec{k}_1 \times \vec{k}_2)_{\parallel}^2}{B_0^2} \frac{mv_m^2}{T_i} \left(\frac{\omega_{k1}\omega_{k2}}{k_{1z}k_{2z}v_m^2}\right)^2 \pi\delta(\omega_{k1} - \omega_{k2} - \omega_{k3})$$
$$\sim \frac{\Omega_i^2}{\delta\omega} \frac{|\delta B_{\omega}|^2}{B_0^2} \int \frac{(\vec{k}_1 \times \vec{k}_2)_{\parallel}^2 \rho_i^4}{(\beta_i + \beta_e)} \frac{\overline{k}_1^2 \overline{k}_2^2}{(1+\overline{k}_1^2)(1+\overline{k}_2^2)} \frac{P(k_{1\perp})}{P_0} dk_{1\perp}. \qquad (38)$$

The integral form of the decay rate given in Eq. (38) was obtained using the same procedure as the integral form of the NL scattering rate given in Eq. (28), as well as Eqs. (3) and (4).

While the three wave decay increase the "number of plasmons" $N_k = W_k/\omega_k$ and "plasmon gas" entropy $\sum \ln N_k$,[19,17] NL scattering by plasma particles conserve $N_k$ $\sum_{k2} \gamma_{NL}^{e,i}(k_1 \to k_2) N_{k2} = Const$. NL scattering by plasma particles as well as three wave interactions are essentially a three-dimensional phenomenon. Two-dimensional simulations of turbulence usually include $B_0$ in the plane of simulation. In such a case, the NL rates due to the dominating non-linear term, that we retained in Eq. (15), disappears in Eq. (28) and Eq. (38) since $(\vec{k}_{1\perp} \times \vec{k}_{2\perp})_\parallel^2 \to 0$. But, we have shown in theory and in simulation, that the essential NL effect can be included in two-dimensional simulations if $B_0$ has two components, in and out of simulation plane.[20]

## 6. A model of KAW turbulence in the solar wind

Comparison of the nonlinear scattering rate Eq. (29) and the three-wave decay rate Eq. (38),

$$\frac{\gamma_{NL}^{k1 \to k2}}{\gamma_{NL}^{3\omega}} \sim \frac{\omega v_m}{v_{tc}} \frac{1}{\bar{k}^6}, \tag{39}$$

shows that for wavelengths, $k^2 c^2 / \omega_{pe}^2 \ll 1$, the NL scattering by plasma electrons dominate over wave-wave interactions in a broadband turbulence. Using observational data[3] (see Fig. 1) we estimate the spectral amplitude as $(\delta B_\omega / B_0)^2 k_\perp P(k_\perp)/P_0 \sim 10^{-3}$ for $k_\perp \rho_i \sim 1$. From Eq. (28) and (38) we estimate the NL wave-particle scattering rate and the three-wave decay/coalescence rate for $k_\perp \rho_i \sim 1$ as $\gamma_{NL}^{k1 \to k2} \sim 10^{-3} \alpha (v_m/v_c)(M/m)^{1/2}(\Omega_i^2/\omega)$ and $\gamma_{NL}^{3\omega} \sim 10^{-3}(\Omega_i^2/\omega)(m/M)^2$. The frequency of the solar wind magnetic fluctuation spectrum is unknown. Under the assumption that $\Omega_i^2/\omega^2 \sim 10$ and using $\Omega_i \sim 1/s$ as found in the in-situ data[3] we get an estimate $\gamma_{NL}^{k1 \to k2} \sim 0.1 \omega \alpha (v_m/v_c)$ and $\gamma_{NL}^{3\omega} \sim 10^{-8} \omega$. Thus $\gamma_{NL}^{3\omega}$ is too small to be important for spectra evolution around $k_\perp \rho_i \sim 1$ because solar wind time of flight to earth $\sim 10^5$ second and NL scattering by electrons is large enough to dominate the turbulence dynamics. To obtain the parameters $\alpha, v_m$, and the ion distribution function as well as turbulent spectra $|\delta B_k|^2$, the evolution of cold and bulk electron and ion distribution functions should be calculated self consistently by balancing the linear and NL landau damping of the KAW with the incoming wave energy flux, $F(\omega/k_\parallel)$, from inertial range as an input parameter in more accurate integral equations which we approximated as Eq. (28) and Eq. (38). This remains as a future task.

From the general integral equation[15-17] for three-wave interaction in weak turbulence theory it follows that the three-wave coalescence has the same rate as decay, $dN_{k3}/dt = \sum_{k2} \gamma_{NL}^{3\omega}(k_1 \to k_2) N_{k2} \delta_{k3-k2-k1}$. Coalescence begins to compete with decay when the number of plasmons $N_{k2} = W_{k2}/\omega_{k2}$ becomes comparable to $N_{k1}$.[19] It is important to note that the induced NL scattering by particles as well as three-wave decay have a threshold which is controlled by the linear wave damping rate, $dN_{k2}/dt = (\gamma_{NL}^{k1 \to k2} + \gamma_{NL}^{\omega1 \to \omega2} + \gamma_{NL}^{3\omega} - 2\gamma_{k2}) N_{k2}$, while three-wave coalescence has no such threshold. Thus the quasi-steady state energy density $|\delta B_{k3}|^2/4\pi$ is defined by equation,

$$2\gamma_{k3}\frac{|\delta B_{k3}|^2}{4\pi\omega_{k3}} = \sum_{k2}\gamma_{NL}^{3\omega}(k_1 \to k_2)\frac{|\delta B_{k2}|^2}{4\pi\omega_{k2}}\delta_{k3-k2-k1}, \qquad (40)$$

where $\gamma_{NL}^{3\omega} \sim \sum|\delta B_{k1}|^2$ is expressed by Eq. (38). The turbulent energy and spectrum in the scattering range where there is no Landau damping, and the level of turbulence at the end of scattering range and the beginning of dissipation range, i.e. at $k_\perp c/\omega_{pe} \sim 1$, are determined by the steady state balance of the turbulent energy influx from the inertial range $F(\omega/k_\parallel)$ and the power of the wave coalescence into dissipation range and linear Landau damping there (see Fig. 3),

$$F = \sum_{k2<1}\gamma_{NL}^{k1\to k2}\frac{|\delta B_{k2}|^2}{4\pi} = \sum_{k3>1}\gamma_{NL}^{3\omega}\frac{|\delta B_{k3}|^2}{4\pi} = \sum_{k3}2\gamma_{k3}\frac{|\delta B_{k3}|^2}{4\pi}. \qquad (41)$$

The NL scattering rate of short wavelength KAW into long wavelength KAW given by the integral Eq. (28), is approximately the same as the rate of the inverse process, $\gamma_{NL}^{k1\to k2} \sim \gamma_{NL}^{k2\to k1}$, if the frequency spectral width of turbulence is large enough, i.e. $\delta\omega/\omega \gg V_A/v_{te}$, which is readily achieved in the steepened solar wind spectra.[3-5] As a result of KAW scattering by particles the energy is quickly redistributed in $k_\perp$-space and a quasi-steady state spectrum $P(k_\perp) \sim |\delta B|^2 \, (\varepsilon^{\nu-1}/k_\perp^\nu)$ established, where $\varepsilon \ll 1$. The scattering power of waves with $k_{1\perp} < k^*$ into $k^*c/\omega_{pe} \sim 1$ where wave coalescence begins to dominate is,

$$Q(k_{1\perp} \to k^*) \sim k^*P(k^*)\gamma_{NL}^{k1\to k^*} \sim k^*P(k^*)\int_\varepsilon^{k^*}dk_{1\perp}(k^* - k_{1\perp})k_{1\perp}k^*P(k_{1\perp}). \qquad (42)$$

The exact solution of Eq. (42) yields the spectral index $\nu$ however the solution is beyond the scope of this article. An approximate solution can be obtained by looking at the nature of the spectrum from Eq. (42). For example, when $\nu = 3$, $Q_{\nu=3} \sim |\delta B|^4 \varepsilon^3(1+\varepsilon \ln \varepsilon)$, while for $\nu = 2$, $Q_{\nu=2} \sim |\delta B|^4 \varepsilon^2(\ln 1/\varepsilon - 1)$. We can conclude that the wave energy flux for $\nu > 3$ is likely too small to transfer the incoming flux $F$ to the dissipation range at $k^*c/\omega_{pe} > 1$, while for $\nu < 2$ the wave energy flux is too large to match the incoming flux $F$ at $k^*c/\omega_{pe} > 1$. Thus the bounds on the spectral index $2 < \nu < 3$ are consistent with the measured spectrum in the solar wind.[3,4] Based on the above arguments the origin of steepening between the inertial and dissipation ranges can be attributed to the nonlinear wave-particle scattering.

## 6. Discussion

The wave-particle interaction of KAW leads to a plateau-like electron distribution function in velocity range $V_A < \omega/k_\parallel < v_m \sim v_{te}$. Inside of interval $|\omega/k_\parallel| < V_A$ there are no KAW with the appropriate phase velocity for linear Landau resonance and the electron distribution remains quasi-Maxwellian, only changing slowly due to NL Landau damping. Outside of interval $|\omega/k_\parallel| > v_m$, KAW Landau damping can take place for $(k_\perp c/\omega_{pe}) > mv_m^2/T_i$ for the model electron distribution function Eq. (13). In the range $\beta_i M/m >> k_\perp^2 \rho_i^2 >> 1$ KAW NL scattering by particles dominate. NL scattering increases or decreases the wave vector $k_\perp$ significantly in accordance with the rate given in Eq. (28), while slightly decreasing frequency with a lower rate given in Eq. (29). The balance of energy is dissipated into the cold electron and ion as heating. In contrast to the claim[1,2] that the collisionless plasma entropy does not change while interacting with waves, in the weak turbulence model the entropy of particles and waves increases.

Measurements of the magnetic field spectrum of solar wind turbulence show that for $k_\perp^2 \rho_i^2 \gg 1$ the magnetic fluctuation $\delta B/B$ is small, and it is possible to expand the Vlasov equation over this small parameter and use the ideas and techniques of weak turbulence theory. It was mentioned by Hasegawa and Chen that NL scattering rate of KAW is larger than the MHD cascade rate by factor $\Omega_i^2/\omega^2$.[9] The rate of KAW NL scattering by particles given by Eq. (26) has an additional large factor $k_\perp^3 \rho_i^3$ which makes it a dominant non-linear process. Additionally there is a possibility of resonant decay/coalescence of KAW into another KAW's. The rate of this decay is smaller than NL scattering by particles by a factor of $(kc/\omega_{pe})^6$.

The global evolution of KAW spectrum in the solar wind can be described as follows and shown in Fig. 3. The wave energy flux $F$ flows from the inertial range $k_\perp \rho_i < 1$ into the scattering range $M\beta_i/m \gg k_\perp^2 \rho_i^2 \gg 1$ where KAW given by Eq. (14) and fills the volume in $k_\perp$ space limited by phase velocity range $V_A < |\omega/k_\parallel| = (v_m^2 + T_i/m)^{1/2} \bar{k}/(1+\bar{k}^2)^{1/2} < v_m$. This $k_\perp$ space volume may be thought of as a tank in which the energy is stored while the KAW plasmons undergo multiple scattering by plasma particles. The wave vector $k_\perp$ increases or decreases (which is not a Kolmogorov-like cascade process) and spectral distribution $|\delta B_k|^2 \sim P(k_\perp) \sim k_\perp^{-\nu}$ is established with $2 < \nu < 3$. Simultaneously KAW frequency decreases by small increments, $(\omega_{k1} - \omega_{k2})/\omega_{k1} \sim (v_{tc}/v_m)(\bar{k}_{2\perp} - \bar{k}_{1\perp})/\bar{k}_{1\perp}\bar{k}_{2\perp}$. The three-wave interaction is important for $\bar{k}_\perp = k_\perp c/\omega_{pe} \sim 1$. As a result of wave coalescence, some fraction of waves are created with larger frequencies and are redistributed in the tank due to NL scattering by particles increasing spectra width over frequencies. The remaining fraction of waves created by waves coalescence falls out of tank into the dissipation range $k_\perp c/\omega_{pe} > 1$, where the electron Landau damping is strong.

A similar examination of the effects of nonlinear scattering by plasma particles for the high frequency $\omega \gg \Omega_i$ magnetosonic/whistler waves in solar wind plasma is now under investigation. Recently we have studied the importance of NL waves-particle scattering in the development of magnetosonic/whistler turbulence in the magnetosphere where Landau damping rate in the background low beta plasma is small.[20] Dispersion relation for the long wavelength $k_\perp c/\omega_{pe} < 1$ electromagnetic magnetosonic/whistler waves $\omega^2 = k^2 V_A^2 (1+\beta) + k_\parallel^2 v_{te}^2 \bar{k}^2/\beta_e$ in a high beta plasma of solar wind is very similar to KAW if $k_\parallel^2/\bar{k}^2 > \beta_e m/M$. However for smaller $k_\parallel^2/k^2$ ratio the phase velocity can be much larger than electron thermal velocity. Preliminary results indicate that in addition to the plateau formation by whistlers, the magnetosonic waves can also lead to a tail formation, which has also been observed in the solar wind.[21] We will report the results subsequently. Thus a careful analysis of the distribution function may be a reliable diagnostic that can determine whether the turbulence is due to kinetic Alfven waves or whistlers/magnetosonic waves or both. It is therefore important to accurately measure the electron distribution function. The current measurements in solar wind plasma do not provide enough resolution of the low energy parts of the distribution and hence the formation of a plateau is not unambiguously visible. On the other hand the tail formation is more clearly visible in the data.[21] Based on our conclusions we recommend that in future experiments an effort be made to accurately measure the low energy portion of the electron distribution function in the solar wind.

## Acknowledgments

This work is supported by the Office of Naval Research. The authors also thank Adolfo Figueroa-Vinas and Olga Alexandrova for valuable discussions concerning solar wind observations.

**Figures :**

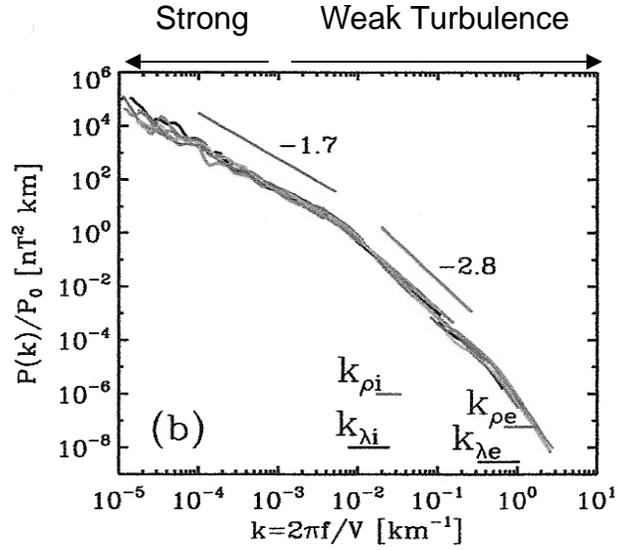

**Figure 1.** The solar wind magnetic field spectrum from Cluster data.[3] Here $k_{\rho i,e} \sim 1/\rho_{i,e}$, $k_{\lambda i,e} \sim \omega_{pi,e}/c$. In the inertial range, $k_\perp \rho_i < 1$, $(\delta B)^2 \sim k^{-5/3}$, and the field amplitude shows an order of magnitude decline where weak turbulence theory can be applied. Between $1 < k_\perp \rho_i < 10$ the steepening may be attributed to NL scattering by electrons which provides an approximate range of $2 < \nu < 3$ for $|\delta B_k|^2 \sim P(k_\perp) \sim k_\perp^{-\nu}$. Finally at $k_\perp c/\omega_{pe} \sim 1$, NL wave coalescence can transfer energy from the scattering range to the dissipation range where electron Landau damping of KAW is strong.

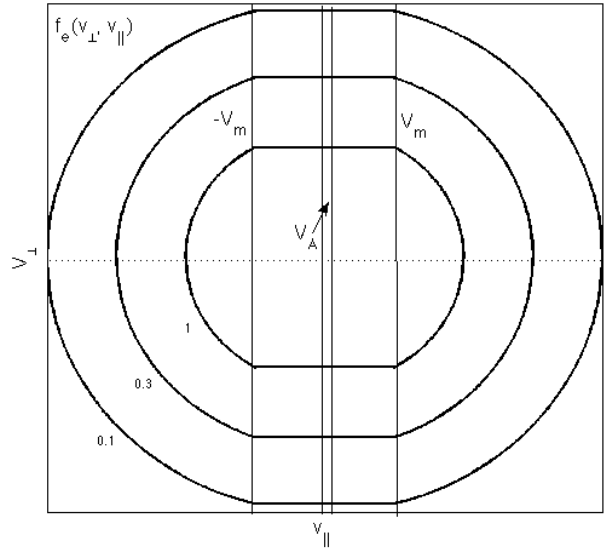

**Figure 2.** The model electron distribution function, Eq. (13). Quasi-linear diffusion establishes and maintains a plateau in electron distribution in the velocity range $V_A < v_\parallel < v_m$, where $v_m$ is less than the bulk electron thermal velocity. Electron Landau damping does not exist here and NL scattering dominates. For the interval $v_\parallel < V_A$, KAW cannot meet the wave-electron phase velocity resonance, and a quasi-Maxwellian distribution remains with density $\alpha n_0$. Outside these regions, the distribution is again Maxwellian.

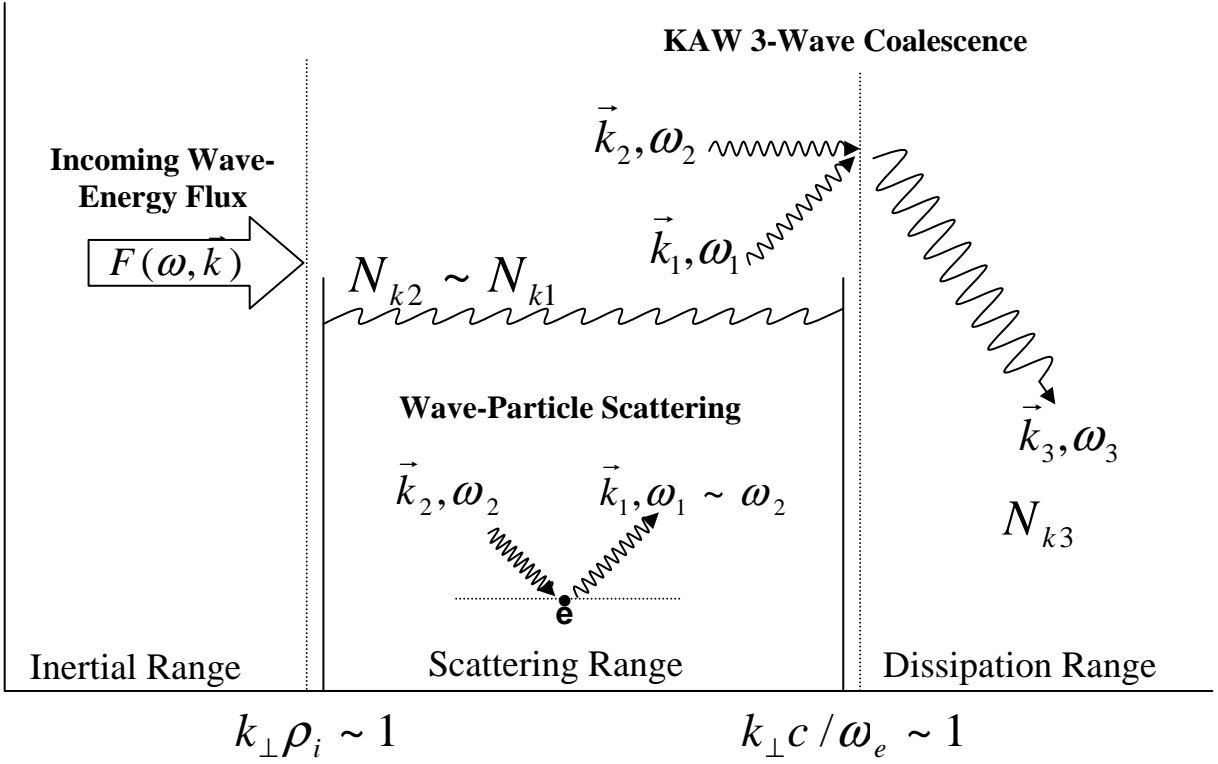

**Figure 3.** The incoming wave energy flux $F(\omega, \vec{k})$ enters from the inertial range into the scattering range where KAW fills the volume in $k_\perp$ space. This volume is like a tank in which the energy is stored while undergoing multiple scattering by plasma particles with increasing or decreasing $k_\perp$. The three-wave interaction rate becomes comparable for $k_\perp c / \omega_{pe} \sim 1$. When the wave amplitude rises such that $N_{k2} \sim N_{k1}$ wave coalescence becomes possible. As a result of coalescence some fraction of waves are created with larger frequencies and are again redistributed in the tank due to NL scattering. The NL scattering and three-wave decay have a threshold which is controlled by the linear wave damping rate, $dN_{k2}/dt = (\gamma_{NL}^{k1 \to k2} + \gamma_{NL}^{\omega 1 \to \omega 2} + \gamma_{NL}^{3\omega} - 2\gamma_{k2})N_{k2}$, while three-wave coalescence has no such threshold. The remaining fraction of waves created by wave coalescence falls out of tank into the dissipation range $k_\perp c / \omega_{pe} > 1$, where they are damped by electron Landau damping.